# Frequent JJ decoupling is the main origin of AC losses in the superconducting state

S. Sarangi[*], S. P. Chockalingam S. V. Bhat

Department of Physics, Indian Institute of Science, Bangalore-560012, India

[*]Corresponding author:

Subhasis Sarangi

Department of Physics

Indian Institute of Science

Bangalore – 560012, India

Tel.: +91-80-22932727, Fax: +91-80-3602602

E-mail: subhasis@physics.iisc.ernet.in

**Abstract:**

The origins of AC losses in the high $T_c$ superconductors are not addressed adequately in literature. We found out, frequent Josephson Junction (JJ) decoupling (both intergranular and the interlayer) due to the flow of AC current is one of the main origins of the AC losses in high $T_c$ superconductors. We have determined the AC losses in superconductors in the rf range by measuring the absolute value of non-resonant rf power absorbed by the samples. Our data shows that under certain conditions when both the number density of JJs present in the sample and the JJ critical current cross a threshold value, AC losses in the superconducting state keeps on increasing with decreasing temperature below $T_c$. The underlying mechanism is an interesting interplay of JJ coupling energy and the amplitude of rf voltage applied to the sample. The effect of an applied magnetic field, variation of rf frequency and temperature were studied in detail. To find out the exact relation between the JJ coupling energy, JJ number density, applied AC frequency, the amplitude of AC current and the AC losses in superconductors, we have studied samples of different crystalline properties, different grain sizes, pressurized with different pressure and sintered at different physical and chemical situations. The implementations of these results are discussed. These results have important implications for the understanding of the origin of AC losses and characterization of superconducting samples. In this paper we also extend the capability of the AC losses studies in superconductors for the characterization of materials for device applications.

Keywords: YBCO; AC losses; rf absorption; Josephson Junction; superconductivity;

**Introduction:**

Zero resistance or zero loss in superconducting materials means that there is no voltage drop along the material when a current is passed through, and by consequence no power is dissipated. This is, however, strictly true only for a DC current of constant value. This is not true in case of AC. The exact reason for the appearance of resistance and the power dissipation in case of AC is not addressed properly in literatures. Hysteresis losses, eddy current losses, self-field losses and the loss due to the flux motion are being treated as the main sources of the AC losses in high $T_c$ superconductors (HTSC) [1, 2, 3, 4]. From the applications point of view, one of the crucial parameter is the power dissipation, especially when subjected to AC fields, which needs to be minimized. Passing AC in superconducting materials without dissipation is one of the major challenges for the various applications of superconductors. So it is very much important to understand and find out all the main origins of AC losses in HTSC. We have studied the AC losses of various superconductors (both in polycrystalline and single crystal) to find out its exact origins in details. We found out, frequent JJ decoupling (both intergranular and interlayer) is one of the most important factors controlling the AC losses in the superconducting samples below $T_c$. JJ decoupling is basically the breaking of a JJ to normal state either by applying current, magnetic field or heat energy. Energy is absorbed or emitted in the formation of annihilation or creation of the Josephson junctions in the prices of JJ decoupling.

Measurement of the exact value of electromagnetic absorption (rf, microwave) by superconducting sample at various physical and chemical situations is a sensitive tool for the detection of AC losses at high frequencies. For both application and fundamental reasons, investigation of the frequency variation of the AC losses, represent one of the important approaches to understand the nature of high $T_c$ superconductors. AC losses have been measured in different polycrystalline $YBa_2Cu_3O_7$ (YBCO) sample at rf frequencies starting from 10 MHz to 20 MHz and temperature 10 K to 300 K. In general it is expected that the AC losses in superconductors should decrease below $T_c$. But surprisingly we found in certain type of superconducting samples, the AC losses suddenly drops very near to zero just below $T_c$ and keep on increases slowly with decreasing temperature below the critical temperature, which property is also strongly dependent on the applied AC frequency and the magnetic field. Investigation of the above phenomena gave us the information regarding the contribution of frequent JJ decoupling towards the AC losses in superconducting samples. Here we explain the loss mechanism with the frequent decoupling of Josephson junctions present in the superconducting samples. Josephson junctions are intrinsically present in all types of superconducting sample. It is found out that even single crystals made with almost all cares are not free from intrinsic JJs or weak links. The number density of JJ depends on the sample type and the sample quality. It also depends on the stoichiometry of sample preparation. The number density of JJ present in granular polycrystalline samples is more than thin film and single crystalline samples. Contacts between the grains in polycrystalline samples and the defects, scratches and the layer structures in the single crystalline samples are the main sources of Josephson junctions. The influence of the frequent Josephson junction decoupling on the AC losses in superconductors is studied in two specially prepared samples where one is full with JJs and the other one is free from JJs.

**Experiment:**

In our experiments we investigated in the high $T_c$ cuprate superconductor YBCO. The YBCO powder used was made by solid-state reaction of reagent-grade $Y_2O_3$, $BaCO_3$ and CuO. We prepared two pellets made of polycrystalline samples of YBCO. The two pellets were exactly similar in size and shape but prepared with different techniques. One of the two pellets (Sample 1) is perfectly sintered and all the precaution has been taken care to minimize the presence of Josephson junctions in it whereas the second pellets (Sample 2) is prepared keeping in mind to make it granular in nature and to enhance the possibilities of Josephson junctions in it. Both the pellets were prepared from the same phase of YBCO powder. First pellet was prepared from the YBCO powder after grinding further for 5 hours and pressurizing with a pressure of 2 tons (sample 2). The second pellet was prepared from the same YBCO powder without grinding further but sintered at temperature of 980 $^0C$ in flowing Oxygen for 8 hours (sample 1). Both the samples show sharp superconducting transition temperature at ~91 K as determined by ρ~$T$ and ac susceptibility measurements. Both the materials were found to be single phasic as determined by x-ray diffraction. The SEM image of the two samples surfaces is shown in Fig. 1. The SEM image of sample 1 shows a much more dense YBCO microstructure than sample 2. This dense microstructure in the sample 1 reduces the number density of JJs present in the sample and also makes the existing JJs stronger. The granular nature in sample 2 is more favorable for weak JJs. So in the sample 2 the number density of JJ is more than sample 1.

The AC losses are calculated from the non-resonant power absorbed by the sample in the presence of rf. The sample is kept inside the rf coil. When we keep the sample inside an rf coil, the rf field generated inside the coil induces rf current in the sample. This rf current flows through the sample. Any loss due to the passing of rf current in sample is compensated by drawing more current from the rf sources. So the AC loss in rf range is determined from the non-resonant power absorbed by the sample. The power absorption measurements were performed with using a standard Integrated Circuit Oscillator (ICO) [5]. This system consists of a self-resonant LC tank circuit of an oscillator driven by a NOT logic gate. The samples under investigation are placed in the core of the coil forming the inductance L and the AC loss is determined from the measured change in the current supplied to the oscillator circuit. The frequency of the oscillator and the rf amplitude were set at 10 MHz and 0.7 Volt for all the temperature variation measurements. Keeping the sample inside the rf coil is equivalent to pass rf current directly in the sample. The sample and the rf probe were arranged inside an Oxford instrument cryostat. No precautions were taken to expel the earth magnetic field. The AC losses measurement was performed on both the sintered and non-sintered pellets (sample 1 and sample 2) having equal size. Rectangular bars with dimensions of $7 \times 3 \times 3$ mm were cut from the sample pellets for the AC losses measurements. Extensive studies have been performed from 10 to 300 K. The AC loss was measured in the magnetic field range of -150 G < $H$ < 150 G. Measurements were made after stabilizing the temperature for about 10 min prior to each reading.

**Results & Discussion:**

Figure 2 shows the AC losses of sample 1 measured at frequency of 10 MHz. The superconducting transition is clearly visible in the figure at the temperature of 91 K. Just below transition temperature sample 1 shows AC losses nearly zero upto the temperature of 13 K. Inset of figure 2 shows the clear picture of the magnetic field dependent AC losses of the sample 1 at 20 K, the loss increases with the magnetic field. The AC loss behavior of sample 2 is just opposite to sample 1 (fig. 3), here instead of decreasing, the AC losses increases below $T_c$ with decreasing temperature. The sample 2 shows AC loss of 400 μ$W$ at the temperature of 13 K and follows an Ambegaokar-Baratoff function for tunnel junction [6] like behavior below $T_c$ (Fig. 3). Inset of figure 3 shows the clear picture of the magnetic field dependent AC losses of sample 2 at 20 K. The phase of magnetic field dependence of sample 2 is just opposite to sample 1. The AC loss decreases with magnetic field in case of sample 2 whereas the loss increases with magnetic field in case of sample 1. At normal state the loss behavior for both the samples follow nearly the same patterns but of different values of AC losses. Sample 1 shows more losses at normal temperature than sample 2, this may be due to the large eddy current loss in sample 1, which is minimized due to the granular nature of the samples 2. But the sample 1 shows the loss, less in superconducting state comparing to the sample 2, which is more important for the various AC applications of superconductors. To know details about the frequency dependence of the AC loss of the sample 2, we have done experiments at three different frequencies at 10 MHz, 15 MHz and 20 MHz respectively (fig. 4). Here it is found that, at all the frequencies, it follows nearly Ambegaokar-Baratoff function below $T_c$. The amplitude of the dependence below $T_c$ increases with frequency. At the temperature just below $T_c$ for all the plots, the behavior does not follow the Ambeogkar-Baratoff function properly. This is discussed in details latter. The behavior of the AC losses with frequency for both the samples is shown in the inset of Fig. 4 at the temperature of 20 K. The AC losses of the granular sample (sample 2) increases with frequency in the range of frequency from 10 MHz to 20 MHz with an average change of 50 X $10^{-12}$ *Watt* / Hz, but for the sintered sample (sample 1) it increases with an average change of only 3 X $10^{-12}$ *Watt* / Hz. The loss in sample 1 is very less comparatively to the sample 2. But in both the cases the fitting is linear which shows the deep interconnection between the AC losses and the frequency of applied AC current.

It is important to understand the differences in the AC losses behavior between the two samples at different rf frequencies and their magnetic field dependences; we have given the following explanation. Between the two superconducting samples, sample 1 is the standard superconducting sample, which shows the AC losses remains very near to zero throughout the superconducting state (Fig. 2). As it is already discussed that sample 2 is a specially prepared sample to enhance the number density of Josephson junctions and the grain boundary weak links. Sample 2 can be model as an array of weakly Josephson-coupled, strongly superconducting anisotropic grains. The intergranular weak links are not only insulating barriers but instead proximity junctions, coupled via semiconducting, normal conducting, or poorly superconducting materials. Due to the occurrence of the large number of weak Josephson junctions in the sample 2, frequent JJ decoupling in the sample 2 is more and due to this, it shows the AC loss more than sample 1 and increases with lowering temperature in the superconducting state (Fig. 2). To know more details, lets explain how these frequent JJ decoupling gives these results; first let us consider, what will happen to a single Josephson junction when we pass current more than its critical value. The obvious answer is it will break the junction and the junction will become normal. The induced rf current in the superconductor mostly flows in the surface of the superconducting sample depending on the London penetration depth $\lambda$. So the rf current flows through out the surface of the sample including the JJs present in the surface. Here we assume that the amount of rf current flowing in the sample wont be able to break all the JJs but only breaks the weaker JJs which have critical current less than the rf current. It is important to note here that the intergranular or the interlayer JJ critical current is very low comparing to the critical current of grain or the layer itself. As earlier discussed each Josephson junction is associated with some coupling energy. This is known as JJ decoupling energy. JJ decoupling energy $E_j$ between neighboring grains is given by [7]

$$E_J(T,H) = \frac{\hbar}{2e} F(T) \langle I_0 \left| \frac{\sin \pi \phi / \phi_0}{\pi \phi_0} \right| \rangle \quad \text{------- (Eq. 1)}$$

Where $\Phi_0 = ch/2e$ is the flux quantum and $F(T)$ is a function of the temperature which in the Ambegaokar-Baratoff theory [6] is given by…

$F(T) = \{\Delta(T)/\Delta(0)\} \tanh \{\Delta(T)/2K_BT\}$ ------------------- (Eq. 2)

with $\Delta(T)$ the temperature dependent gap parameter and $\Delta(0)$ the gap at $T = 0$; $I_0$ is the maximum Josephson current given by

$$I_0 = \frac{\pi \Lambda(0)}{eR_n} \quad \text{---------------- (Eq. 3)}$$

with $R_n$ the normal state resistance of the junction; $\Phi = HA_J$ and $A_J$ is the effective field penetration junction area orthogonal to $H$. In the average Eq. (1) refers to the statistical distribution of the junction geometrical parameters. So once the Josephson junction breaks, the system will absorb energy $E_j$, this is a function of both the temperature and magnetic field. To make the experiment simple, we have done all the temperature dependent AC losses measurements at zero fields. This eliminates the magnetic field $H$ contribution in the above equation. The exactly opposite situation will happen when we reduce the current across the critical current. The JJ, which was decoupled before by the application of higher current, will come back to its original situation and the system will release exactly same amount of energy $E_j$ just below $J_c$ and will become a Josephson junction again. If we pass an AC current in the Josephson junction having amplitude more than the critical current of Josephson junction then the system will absorb and release energy continuously as represented in the figure 5. At the point 'a' the current passes through the junction is more than its critical value, so the junction will absorb energy and at the point 'b' it will emit the energy in the form of heat. So the expression for the total energy absorbed by the sample per second or the total loss can be expressed as $P$ and $P = 2fNE_j$ (Eq. 4), where $f$ is the frequency of the AC current, $N$ is the total number of Josephson junctions keeps on breaking and forming due to the application of AC current and $E_j$ is the JJ decoupling energy of a single JJ.

The total energy absorbed by the sample due to the passing of rf current is a measure of loss and can be treated as the AC losses of the system at frequency f. From this explanation (see Eq. 4) it is understood that if the sample have more JJs or higher the JJ critical current (not more than the applied AC current because in that case the JJ does not decouple) then the loss will be more. From the Fig. 1 it is clear that the sample 2 has more number of JJs present than the sample 1, so it shows AC losses more in the superconducting state and due to the presence of very less number of JJs in sample 1, it shows the AC losses very near to zero at the same range of frequency. Now to explain the Ambeogkar-Baratoff (AB) pattern of the AC loss in the sample 2, we have to look at the equation 2. From the equation 2 it is clear that the critical current of a JJ and the $E_j$ both follows the AB pattern. So the AC loss, which is a linear function of the JJ decoupling energy $E_j$ (equ. 4), has to follow the AB pattern. Just to explain the temperature dependent AC losses of the granular YBCO sample (sample 2) at various frequencies (Fig. 4), we have to see the equation 4 again. From the equation 4 it is clear that the AC losses in the superconducting sample is directly proportional to the frequency of applied AC current. Increasing frequency makes the total number of JJ decoupling per second more so the AC loss due to the JJ decoupling increases linearly with the applied frequency. The application of a magnetic field and the variation of temperature alter the AC penetration depth of the sample, which in turn changes the AC losses associated with it. This is the reason why the magnetic field dependent loss in case of sample 1 increases with field. In the case of sample 2 the frequent JJ decoupling is the dominating factor. The critical current term in the equation 1 decreases with field, so the decoupling energy $E_j$ also decreases with field. Due to the decreasing of $E_j$ with increasing magnetic field, AC loss in case of sample 2 decreases with field.

Here in the above discussion we have assumed all the JJs to be have equal critical current but in the real picture, there will be a distribution of JJ critical current for the whole JJs and that may be one of the reasons why the experimental curves discussed above do not fit exactly with the Ambeogkar-Baratoff function. The Ambeogkar-Baratoff function is applicable for a single Josepson junction. It is possible to give a model considering the whole JJs network in the sample and its corresponding fit and the fit may exactly fit with the experimental data. But the work is in progress and will be discussed in a forthcoming publication [8].

**Conclusions:**

We have intentionally varied microstructures of polycrystalline YBCO samples. We have observed close correlation between AC losses and the microstructures of the superconducting samples. In the view of the discussion made in the previous section it is obvious that the result of the study of the AC losses in the superconductor YBCO by non-resonant rf power absorption are consistent. In conclusion, we have observed a well-defined increase in the AC losses in the granular superconductor with decreasing temperature. The AC losses response obtained below critical temperature in the granular superconducting sample is mainly due to the frequent JJ decoupling. This effect is also seen in other superconductors like BSCCO, LSCO and $MgB_2$ superconducting samples both in polycrystalline and single crystal forms. As the AC losses due to the frequent JJ decoupling are directly proportional to the applied AC frequency, at microwave frequency, it is expected to give bigger contribution. These studies give both quantitative and qualitative knowledge to understand the AC losses in superconductors and may be useful for various AC applications of high $T_c$ superconductors.

**Acknowledgements:**

This work is supported by the Department of Science and Technology, University Grants Commission and the Council of Scientific and Industrial Research, Government of India.

**Figure Captions:**

1. SEM images of the surfaces of both the samples (sample 1, YBCO sintered sample and sample 2, YBCO granular sample). Both the samples show different microstructures due to the different preparation techniques.

2. Temperature dependence of the AC losses for the sample labeled as sample 1 (YBCO sintered sample) at the frequency of 10 MHz. Solid curve is the Ambegaokar-Baratoff fit as explained in the text. The inset shows both the forward and reverse scan for the magnetic field dependent loss of the same sample at 20 K in between −150 G to 150 Gauss.

3. Temperature dependence of the AC losses for the sample 2 (YBCO granular sample) at frequencies of 10 MHz. Solid curve is the respective Ambegaokar-Baratoff fits $F(T)$ [(Eq. 2)] with delta (T) given by the BCS theory, as explained in the text. We interpret the increase in AC losses with lowering temperature is due to the increase of JJ critical current of the junctions. The inset shows both the forward and reverse scan for the magnetic field dependent loss of the same sample at 20 K in between −150 Gauss to 150 Gauss.

4. Figure shows the AC losses of sample 2 at different rf frequencies below $T_c$. Solid curves are the respective Ambegaokar-Baratoff fits. Inset shows the frequency ($f$) dependence of the AC losses for both the samples (sample 1 and sample 2) at $T = 20$ K and 20 MHz $> f >$ 10 MHz. Solid curve is linear fit with different slope, as explained in the text.

5. Picture showing the position of JJ decoupling due the excessive current. $J_c$ is the critical current of the Josephson junction. The rectangular box is the voltage-time area where the JJ exists. The arrow indicates the exact position at which JJ annihilation and formation occurs. Energy equal to $E_j$ is absorbed by the Josephson junction at the point "a" and the same amount of energy is released at the point "b". The energy released at the point "b" dissipates to the sample in the form of heat energy. This occurs twice at every cycle of AC current. One full cycle of AC gives two annihilations and two creations of the Josephson junction. The output voltage is the AC voltage applied to the sample. The figure is made for 10 MHz frequency but applicable for other frequencies.

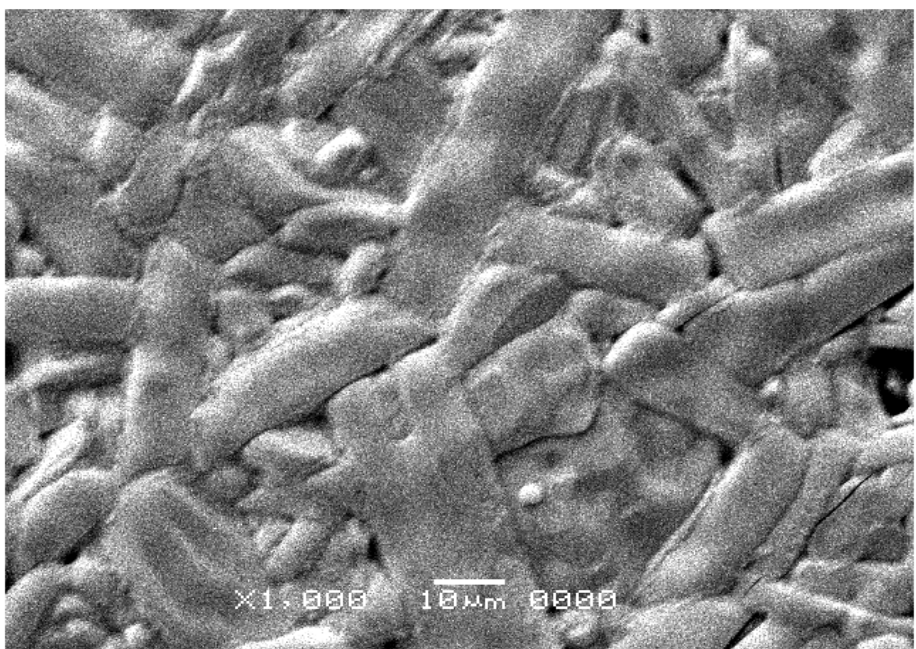

(Sample 1)

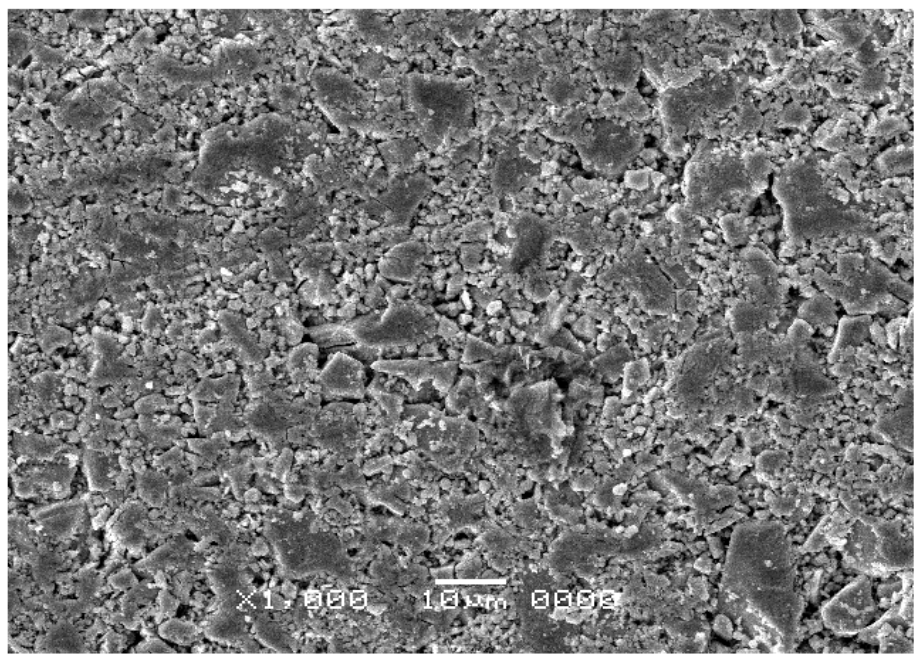

(Sample 2)

**FIG. 1**

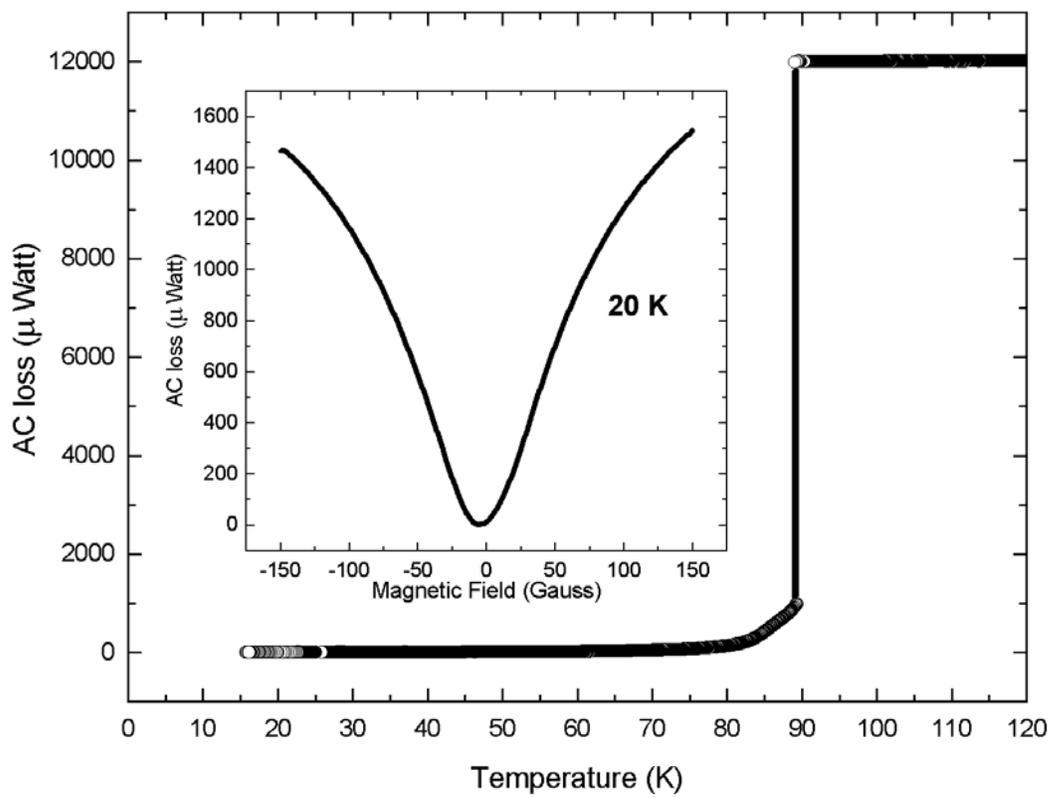

**FIG. 2.**

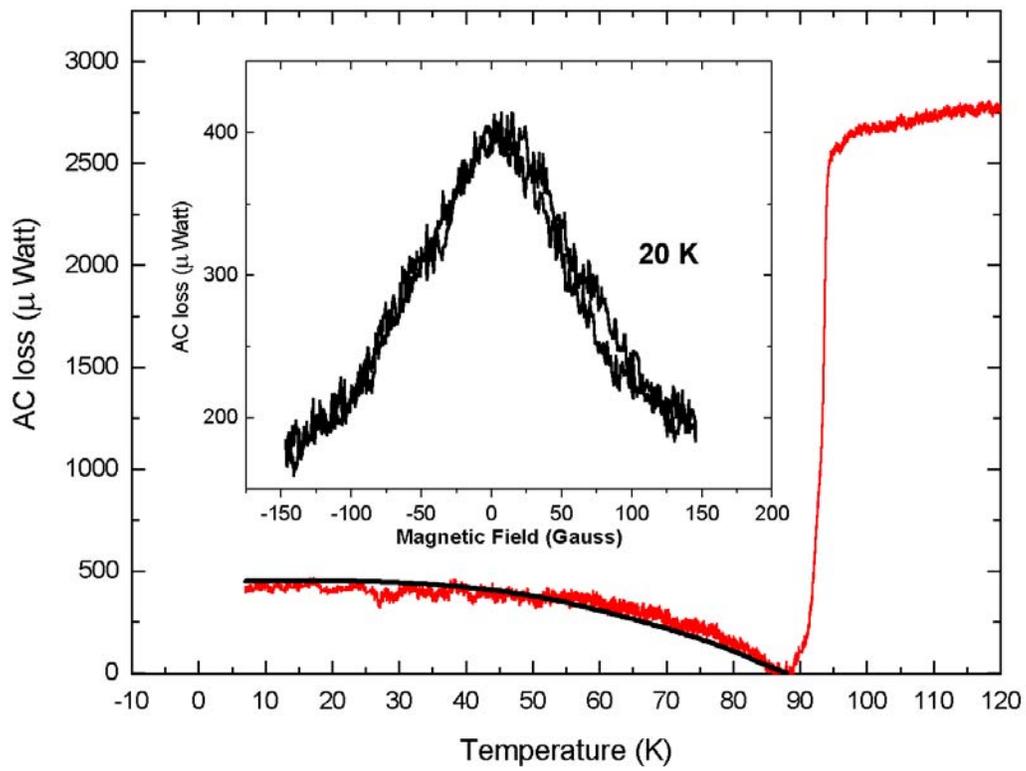

**FIG. 3.**

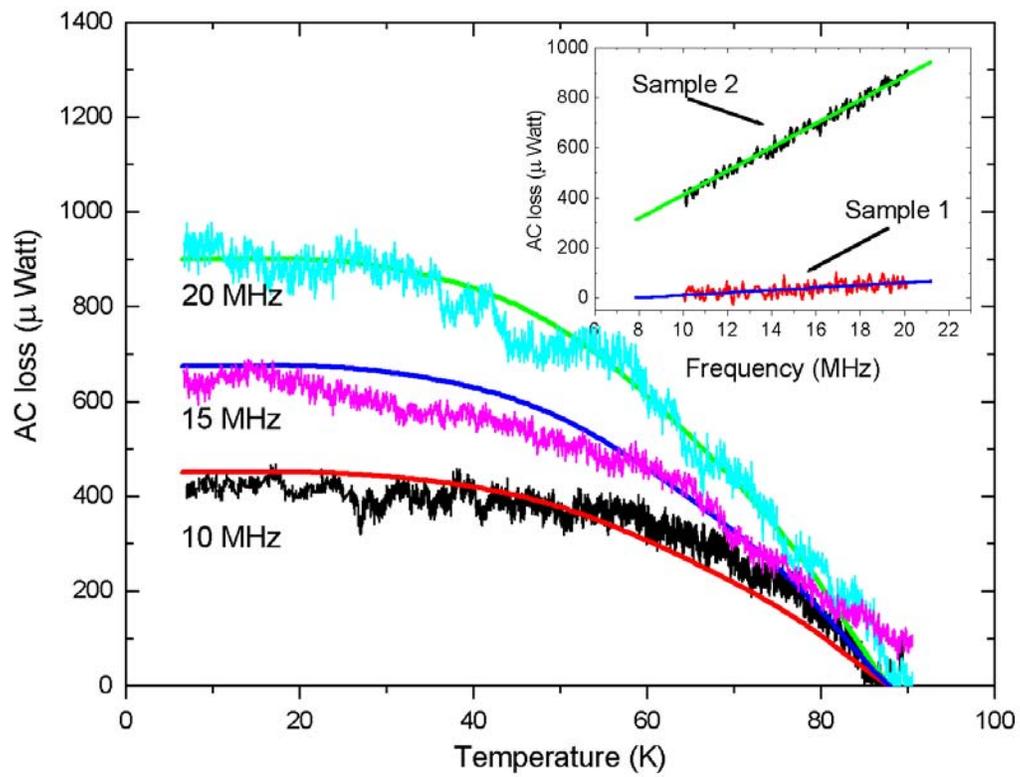

**FIG. 4.**

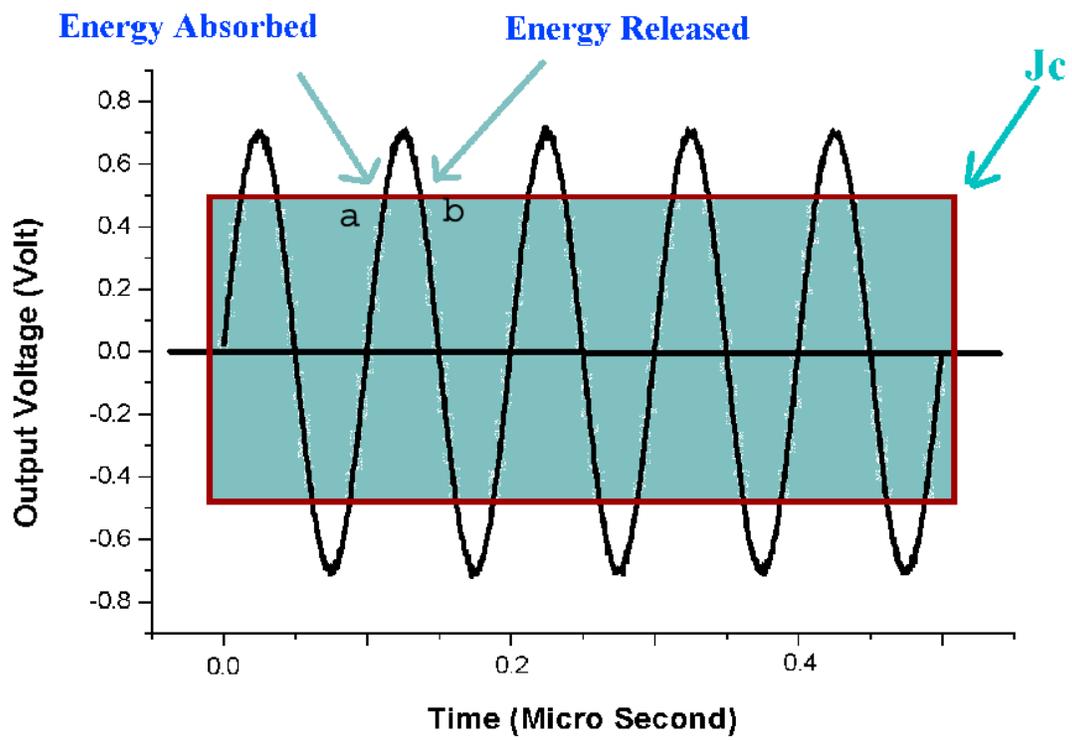

**FIG. 5.**